\documentclass[aps,showpacs,twocolumn]{revtex4}

\usepackage{epsfig}

\newcommand{\beq}{\begin{equation}}
\newcommand{\eeq}{\end{equation}}
\newcommand{\bea}{\begin{eqnarray}}
\newcommand{\eea}{\end{eqnarray}}
\newcommand{\nn}{\nonumber}

\newcommand{\eps}{\epsilon}
\newcommand{\veps}{\varepsilon}

\newcommand{\s}{\sigma}

\newcommand{\be}{\beta}

\newcommand{\vp}{\varphi}
\newcommand{\ra}{\rangle}

\newcommand{\ga}{\gamma}

\newcommand{\ua}{\uparrow}
\newcommand{\da}{\downarrow}

\newcommand{\tht}{\theta}

\begin{document}

\title{Electronic spin precession and interferometry from spin-orbital entanglement in a double quantum dot}
\author{P. Simon$^{a)}$ and D. Feinberg$^{b)}$}
\affiliation{$^{a)}$ Laboratoire de Physique et Mod\'elisation des  Milieux
    Condens\'es, CNRS and Universit\'e Joseph Fourier, BP 166, 38042
Grenoble, France}
\affiliation{$^{b)}$ Institut N\'eel, CNRS, associated to Universit\'e Joseph Fourier, BP 166,  38042
Grenoble, France}

\date{\today}
\begin{abstract}
A double quantum dot inserted in parallel between two metallic leads can entangle the 
electron spin with the orbital (dot index) degree of freedom. An Aharonov-Bohm orbital phase 
can be transferred to the spinor wavefunction, providing a geometrical control of the 
spin precession around a fixed magnetic field. A fully coherent behavior 
occurs in a mixed orbital/spin Kondo regime. 
Evidence for the spin precession can be obtained, either using spin-polarized metallic leads 
or by placing the double dot in one branch of a metallic loop. 
\end{abstract}


\maketitle

Control of the electron spin is important for the realization of novel 
nanoelectronic devices for spintronics or for quantum information processing. In 
the latter case, manipulation of individual spins is necessary. The use of time-dependent gates, 
put forward some years ago \cite{loss-divincenzo}, has seen considerable progress 
\cite{expt-spinqubits}. Yet another possibility is to build 
single or two-spin operations (gates) into a given device geometry, with static control parameters only. 
This may allow faster processing speed and facilitate integration into more complex devices. 
One way of controlling the 
spin in quantum dots is through energy filtering by applied gate voltages, as proposed for 
spin entanglement \cite{entangler}, teleportation \cite{teleportation}, and 
spin filtering \cite{spin-filter}. Spin precession has also been  put forward in metallic rings, due to spin-orbit (Rashba) interaction \cite{precession}. 

In the present Letter, we explore means of achieving individual spin precession in quantum dots, 
which allows fully transparent operation (unitary transmission). It also paves the way for  fundamental 
tests of quantum mechanics which have not yet been realized with single electrons. The basic unit is a 
double quantum dot in parallel, coupled to metallic leads. 
As previously shown in \cite{apl}, combining a strong magnetic field with well-chosen gate voltages allows 
a Zeeman splitting in each dot, such that spins "up" travel through dot $1$ while spins "down" 
travel through dot $2$. The splitting of a non-polarized incoming current into two oppositely polarized 
currents was proposed by us in a three-terminal geometry as an electronic Stern-Gerlach splitter \cite{apl}. 
Here we show that closing the set-up in a loop enclosing an Aharonov-Bohm (AB) flux $\phi$ (
Figure 1) gives rise to new physics. 
Indeed, an incoming state $a|\ua\ra + b|\da\ra$ becomes $a|1\ra\bigotimes|\ua\ra +
be^{i\vp}|2\ra\bigotimes|\da\ra$, then $a|\ua\ra + be^{i\vp}|\da\ra$ at the output
where $\vp=2\pi\frac{e}{h}\phi$ is 
the AB phase between the two branches.
The intermediate state 
involves entanglement between spin and orbital degrees of freedom, thus applying an orbital phase 
causes a rotation of the spinor around the fixed magnetic field. This permits a purely geometrical control of 
the spin precession angle, by static parameters such as the gate voltages and the AB flux. This set-up  indeed belongs to the class of "Stern-Gerlach interferometers", first considered as gedanken experiments \cite{SGI}, then realized with neutron interferometry \cite{SGI-expt}. 
The success of spin precession relies on quantum coherence between the two branches, and no "which-path" information being gained. Thus our proposal based on a nanoelectronic device  provides a 
very sensitive test of decoherence effects.

\begin{figure}[h]
\label{Fig:interfero}
\epsfig{figure=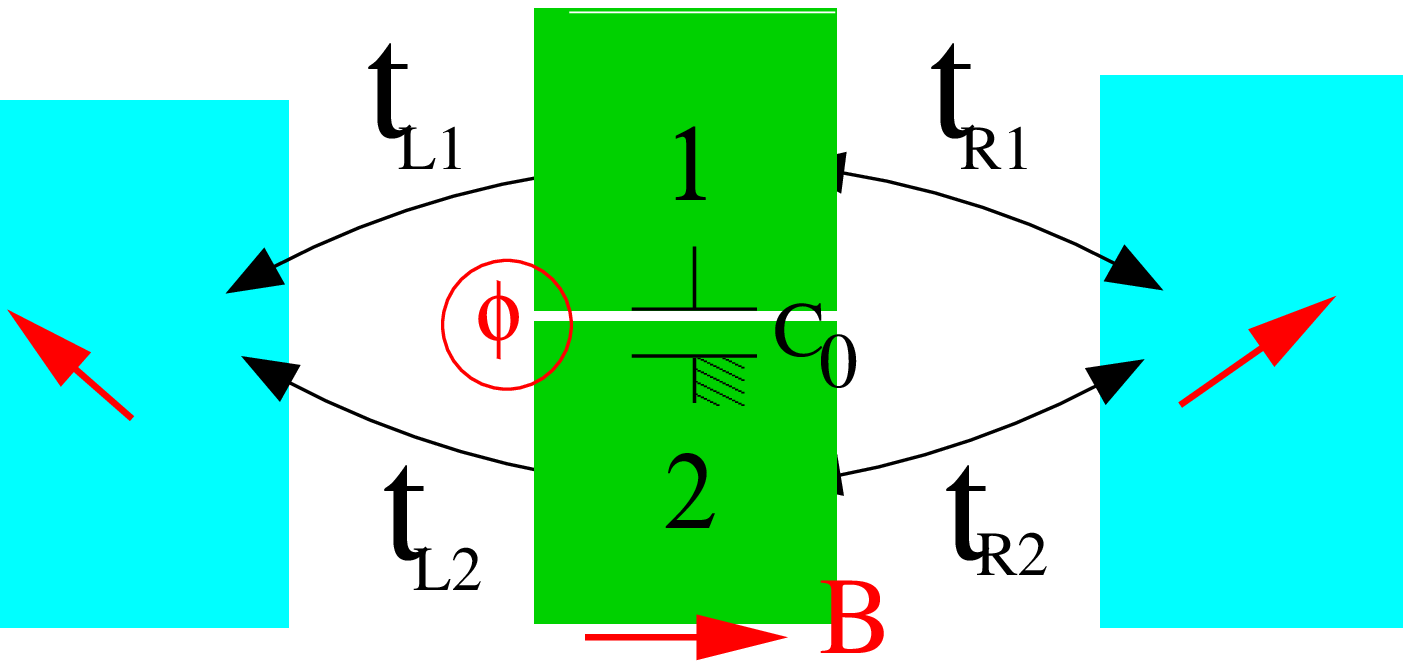,width=5.5cm,height=3.cm,angle=-0}
\epsfig{figure=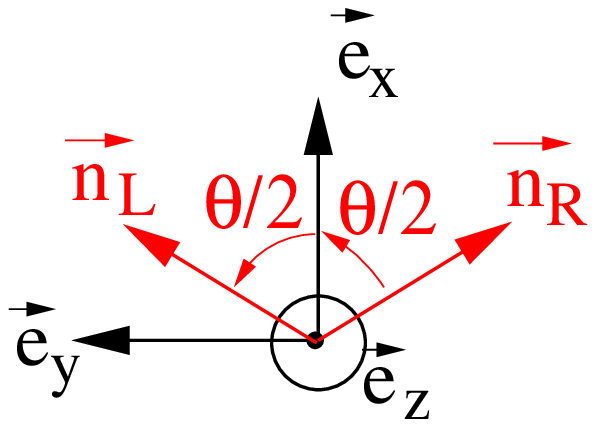,width=2.cm,height=2.cm,angle=-0}
\caption{Left: Schematic representation of the proposed setup:
two small quantum dots coupled by a capacitance $C_0$ and connected to
left and right partially polarized reservoirs. Depending on the choice of
gate voltages,  the upper branch filters spins up and the lower one spins
down, or vice versa. A magnetic flux $\phi$ threads the whole device.
Right: The polarization axis $\vec n_L$ and $\vec n_R$ make an
angle $\tht/2$ with the $\vec e_x$ axis and $\vec B= B \vec e_z$.}
\end{figure}

Let us consider two small quantum dots in parallel, keeping only one orbital level, 
with charge number states $0,1,2$  in each of them. Under an applied Zeeman field, 
it has been shown \cite{spin-filter} that a single dot may filter 
spins "up" or "down", depending on the applied gate voltage $V_{g \nu}$ ($\nu=1,2$) via
the capacitances $C_{g \nu}$. Indeed, in a resonant regime, transitions  between 
number states $0,1$ involve spins $\ua$ only (states $0,\ua$), 
while transitions between number states $1,2$ instead involve spins $\da$ only (states $\ua,\ua\da$). We assume the two dots to be coupled by a large capacitance $C_0$, and label the double-dot 
(DD) states as ($\Psi_1,\Psi_2$). 
The gate voltages are chosen such that the lowest-lying states be $(\ua,\ua)$ and $(0,\ua\da)$ and be
degenerate. Single-electron transitions from (to) the leads involve higher-energy states such as 
$(0,\ua)$, $(\ua,\ua\da)$. One can achieve a Kondo regime where the resonance between states 
$(\ua,\ua)$ and $(0,\ua\da)$ involves spin-up transitions in dot $1$ and spin-down transitions 
in dot $2$. This is an orbital/spin Kondo effect entangling spin and orbital degrees of freedom. 
It achieves a novel sector of the Kondo physics, completing 
the already existing pure spin Kondo \cite{spin-Kondo} and pure orbital Kondo \cite{orbital-Kondo,expt} ones. 
Here, both dots are connected to the same left and right lead with tunneling  hopping
parameters $t_{L \nu}$ and $t_{R \nu}$. In view of a direct detection 
of the spin precession, we allow for spin-polarization in the leads. 
The corresponding tunnel junctions capacitances are  $C_{L}$
(left), $C_{R}$ (right). A magnetic field $B$, oriented in the plane of the setup,
is applied to each dot. The optimum Kondo regime
is reached at the symmetric point where the lowest-lying excited states  
$(0,\ua)$, $(\ua,\ua\da)$ involve the same charge excitation energy
$E_c=\frac{e^2 C_0}{2C(C+2C_0)}$ where $C=C_L + C_R + C_g$ (we assume
$C_g=C_{g1}\approx C_{g2}$). The isolated DD system may be
described \cite{borda} at low energy by 
\beq
H_{dot}=-\delta E T^z-t T^x-g\mu_B (B_1 S^z_1+ B_2 S^z_2),
\eeq
where we have defined the orbital pseudospin  $T^z=(n_1-n_2+1)/2=\pm
1/2$ from the charge occupations $n_{\nu}$.
Here $\delta E=\frac{e}{C+2C_0}[C_g(V_{g2}-V_{g1})-e]$. The second term in $H_{dot}$ represents a
small parasitic tunneling amplitude between the dots \cite{apl}. The last term
expresses the effective Zeeman splitting. Due to exchange contributions with the leads \cite{konig}, the 
local fields $B_{\nu}$ in the dots may be different. Notice that a large level spacing
$\delta\eps$ (or equivalently
small quantum dots) is  necessary
 to eliminate the triplet states  $(0,t)$ \cite{spin-filter}. Under the condition $\delta E=\frac{1}{2}g\mu_B
(B_{1}+B_{2})$, the states $(\uparrow,\uparrow)$
and $(0,\uparrow\downarrow)$ are degenerate. The total spin
$S^z=S^z_1+S^z_2-\frac{1}{2}$ is entangled with the orbital pseudospin $T^z$,
e.g. a  spin flip is locked to an orbital pseudo-spin flip.
Therefore the Kondo screening of the spin involves spin-up electrons in
branch $1$ and spin-down electrons in  branch $2$. Notice that  the gate voltage difference 
compensates for the Zeeman splitting between spins up in dot $1$ and down in dot $2$,
and no splitting of the Kondo zero-bias conductance peak occurs.

The system Hamiltonian is $H=H_{leads}+H_{tun}+H_{dot}$.
The leads are described by
$H_{\rm leads}=\sum\limits_{k,\ga,\pm}\veps_{k\pm} c^\dag_{k,\ga,\pm}
c_{k,\ga,\pm}$,
where $ c^\dag_{k,\ga,\pm}$ creates an electron with energy
$\veps_{k\pm}$ in lead
$\ga=L,R$ with  spin along $\pm \vec n_\ga$. Spin polarization in the leads would result in
 a spin asymmetry
in the density of states $\rho_{\ga\pm}(\omega)$. We further neglect the
energy dependence
in the density of states and also suppose $\rho_{\ga\pm}\approx \rho_{\pm}$.
The ratio $p=(\rho_{+}-\rho_{-})/(\rho_{+}+\rho_{-})$ denotes the degree of
spin polarization in the leads, which may have a noncollinear
polarization.
The tunneling junctions between the leads and the dots are described  by
$H_{tun}=\sum\limits_{k,\ga,s=\pm,\nu} (t_{\ga \nu}  c^\dag_{k,\ga,s}
d_{\nu,s}+H.c.)$ where $d_{\nu,s}$  
destroys an electron in dot $\nu=1,2$ with spin $s=\pm$.

In addition to the applied field, the polarized electrodes may also generate effective magnetic
fields $B_{eff,\nu}$. They depend on $p$, on the dot internal parameters and gate
voltages, and on the tunnel-coupling strength
$\Gamma_{\nu}=\Gamma_{L\nu}+\Gamma_{R\nu}$ where
$\Gamma_{\ga \nu}=\pi \sum_{k s=\pm} |t_{k\ga \nu}|^2 \rho_{s}$.
An explicit mean-field calculation of $B_{eff,\nu}$ is derived in
\cite{konig} where it was shown that $B_{eff,\nu}$ is zero
for a particle-hole symmetric situation. In what follows, we assume that the fields $B_{eff,\nu}$ can be neglected.
We suppose that the lead
magnetization axis lies in the dot plane $(\vec e_x,\vec e_y)$, with a relative angle $\tht$ (see Fig. 1).
Following \cite{konig}, it is convenient to quantize the dot spin
along the B axis corresponding to $\vec e_z=(\vec n_L\times \vec n_R)$, the
other vector coordinates being $e_x=(\vec n_L+\vec n_R)/|\vec n_L+\vec n_R|$,
$e_y=(\vec n_L-\vec n_R)/|\vec n_L-\vec n_R|$.
In this rotated basis, the tunneling Hamiltonian then reads\cite{konig}
\bea
H_{tun}&=&\sum\limits_{k,\nu} \left(t_{L \nu}\;e^{i(-1)^{\nu} \vp/4}
c^\dag_{k,L,\s}d_{\nu,\s}+H.c.  \right)\\
&+&\left(L\to R,\vp\to -\vp\right).\nn
\eea
The $c^\dag_{k,L,\pm}$ are linear combinations of the $c^\dag_{k,L,\ua/\da}$ and are related by
$(c^\dag_{k,L/R,+},c^\dag_{k,L/R,-})=(c^\dag_{k,L/R,\ua},c^\dag_{k,L/R,\da})\,e^{\pm i\tht\s^z/4}\,\frac{(\s^x+\s^z)}{\sqrt{2}}$.
In order to determine the effective coupling between the double dot and  the
leads,
we consider virtual excitation to both excited states $(0,\ua)$ and  $(\ua,\ua\da)$
generated by $H_{tun}$. Using a Schrieffer-Wolf (SW)
transformation, the Kondo Hamiltonian $H_K$ is obtained:
\bea\label{hk1}
H_{K}&=&\frac{1}{2}T^{-}(J_{LL}^\perp  e^{-i\vp/2}\psi^\dag_{L\ua}\psi_{L\da}
+J_{RR}^\perp e^{i\vp/2}\psi^\dag_{R\ua}\psi_{R\da})\nn\\
&+&\frac{1}{2} T^{-}(J_{LR}^\perp\psi^\dag_{L\ua}\psi_{R\da}+
J_{RL}^\perp\psi^\dag_{R\ua}\psi_{L\da})+H.c.\nn\\
&+&\frac{1}{2}T^z(J_{LL}^{z\ua}\psi^\dag_{L\ua}\psi_{L\ua}-
J_{LL}^{z\da}\psi^\dag_{L\da}\psi_{L\da})\\
&+&
\frac{1}{2}T^z(J_{RR}^{z\ua}\psi^\dag_{R\ua}\psi_{R\ua}-
J_{RR}^{z\da}\psi^\dag_{R\da}\psi_{R\da}) \nn \\
&+&  \frac{1}{2}T^z\left(e^{-i\vp/2}
J_{LR}^{z\ua}\psi^\dag_{L\ua}\psi_{R\ua}-e^{i\vp/2}J_{LR}^{z\da}\psi^\dag_{L\da}\psi_{R\da}
+H.c.\right),\nn
\eea
where $\psi_{L/R,\s}=\sum_k c_{k,L/R,\s}$ and
$T^-=d^{\dag}_{2\da}d_{1\ua}$ flips both
the spin and the orbital pseudo-spin.
We have introduced several Kondo  couplings 
$J_{\beta,\gamma}^\perp\approx \frac{t_{\beta,1}t_{\gamma,2}}{E_C}$ and 
$ J_{\beta,\gamma}^{z,\ua/\da}\approx  
\frac{t_{\beta,1/2}t_{\gamma,1/2}}{E_C}$ with $\beta,\ga=L,R$.
As usual the $k$-dependence of
the Kondo couplings is neglected.
The $J_{\beta\gamma}^z$ Kondo couplings are in general spin-dependent, due to intrinsic asymmetries in the 
two branches, except when
$t_{\ga 1}\sim t_{\ga 2}\sim t_\ga$. 
However, due to the entanglement
of orbital and spin degrees of freedom, a geometrical asymmetry is easily compensated
 with an  orbital
field, {\it i.e.} with a fine-tuning of the dot gate voltages  $V_{g1}$ and
$V_{g2}$. If not stated, $t_{\ga 1}\sim t_{\ga 2}\sim  t_\ga$ is therefore
assumed in the following. Other cotunneling terms  involving
higher energy states like $(\ua\da,\ua)$,
turn out to be irrelevant under renormalization group (RG) in the low-energy limit
and thus do not spoil the spin filtering.
In the unitary limit, the spin $\vec S$ or equivalently the pseudo-spin  $\vec T$ is completely screened and an entangled spin/orbital singlet  is formed
with the left and right electrodes. We can get rid of the  phase in the
Hamiltonian in eq. (\ref{hk1}) by defining the new rotated basis
$(\tilde \psi^\dag_{L/R\ua}\tilde\psi^\dag_{L/R\da})=(\psi_{L/R\ua}^\dag\psi^\dag_{L/R\da}){\cal R}^z(\pm\vp/2)$.
In this spin-rotated  basis, the Kondo Hamiltonian takes the
simpler form:
$
H_K=\sum\limits_{\be,\ga}\sum\limits_{s,s'=\pm} J_{\be\ga}\tilde  \psi_{\be
s}^\dag\vec T\cdot {\vec \tau}_{s s'}\tilde \psi_{\ga s'}.
$
The AB phase $\vp$ has disappeared from the the Kondo Hamiltonian and  has been
swapped onto the spin direction in the source and lead. In this basis the angles 
$\tht$ and $\vp$ therefore play a similar role \cite{equiv}. 

For weak polarization, the Kondo temperature is well approximated by  
$T_K\approx\Delta\exp(-\frac{1}{\rho(J_{LL}+J_{RR})})$ with $\Delta\approx
\delta\eps$ the  dot level spacing \cite{cornaglia}.
At $T\ll T_K$, the system reaches the unitary limit \cite{spin-Kondo} and  the
$T=0$ transmission amplitude ${\cal T}_\s$
reads:
\beq
{\cal
T}_\s(\eps_F)=2i\frac{t_Lt_R}{t_L^2+t_R^2}e^{i\delta_\s(\eps_F)}\sin(\delta_\s(\eps_F)),
\eeq
where $\delta_\ua(\eps_F)= \delta_\da(\eps_F)=\delta\sim\pi/2$ is the  Kondo
phase shift and $\eps_F$ the Fermi energy.
Assume one prepares an incoming state $|\eta\rangle_{in}$ in a  superposition
of $\ua$ and $\da$ states,
$|\eta\rangle_{in}=a |\ua\rangle_{L}+b|\da\rangle_{L})$,
the outgoing state reads
\beq
|\eta\rangle_{out}  =2i\frac{t_Lt_R}{t_L^2+t_R^2}e^{i(\delta-\vp/2)}
\left( a|\ua\rangle_{R}+e^{i\vp}b |\da\rangle_{R}\right).
\eeq
One sees that by manipulating the AB flux, one can perform a coherent
precession of the spin of the
incoming wave function around the magnetic field axis.
This effect has a signature in the low-temperature conductance
through the DD:
\beq\label{conduc1}
G_1=G_0\left(1-\frac{2p^2}{1+p^2}\sin^2\frac{(\tht-\vp)}{2}\right),
\eeq
where
$G_0=\frac{e^2}{2h}\frac{4\Gamma_L\Gamma_R}{(\Gamma_L+\Gamma_R)^2}(1+p^2)$ is
the maximum conductance obtained
for parallel lead polarization.
In Eq. (\ref{conduc1}), the conductance reaches its maximum value
for $\vp=\tht$, showing that a small
AB flux compensates the mismatch of non-collinear polarizations.

The above results were derived assuming $T=0$ and
neglecting environmental fluctuations inherent to any mesoscopic setup.
While the DD operates close to the unitary regime $T\ll  T_K$,
inelastic processes induced by a finite $T$
are rather small and their main effects are to reduce the amplitude of  the
outgoing wave vector
and give a dephasing time $\tau_\phi^{-1}\sim T^2$ \cite{nozieres}.
Another source of decoherence is brought by the circuit  electromagnetic
fluctuations, that couple to tunneling events to and from  each of the dots. 
We have used an equivalent circuit  representation of our Stern-Gerlach interferometer (SGI) following Ref.
\cite{IN}. By modeling the environment by an external impedance $Z$, we have shown 
using a RG analysis that only a large impedance
of order $R_K$ is able to destroy the global coherence of the electronic SGI.
This will be detailed elsewhere \cite{simon06}.

The above proposal of detecting the precession directly by spin filtering the source 
and drain electrodes 
is similar to polarized neutron interferometry experiments \cite{SGI-expt}. 
Still, it might be difficult to control the spin filtering in the leads, as high Zeeman fields are necessary for the orbital/spin Kondo effect. These fields can be estimated as 
$B \sim $ few $T$ for GaAs-based dots, and one order of magnitude less for InAs dots owing to the much larger g-factors. 

\begin{figure}[h]
\label{Fig:doubleinterfero}
\epsfig{figure=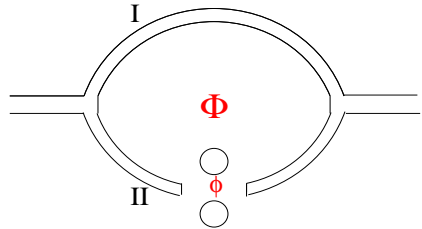,width=5.5cm,height=3.cm,angle=-0}
\caption{
The double dot is embedded into a larger AB interferometer.
Two AB flux $\Phi$ and $\phi$, corresponding to the large and small loops, have been introduced.}
\end{figure}

Alternatively, 
there exists another way of detecting the precession. 
Suppose one inserts the DD in the branch (II) of a larger loop including a flux $\Phi$ \cite{choi00}(Figure 2). 
The whole loop is supposed to be phase-coherent. The transmission through the upper branch (I), as well as the
scattering matrices at the loop extremities do not involve any spin dependence. 
AB interference in the large loop 
amounts to adding a spin-conserving amplitude in branch (I) and a spin-precessing amplitude in branch (II). 

Consider first for simplicity an open AB interferometer (ABI) realizing the equivalent of 
Young's double-slit 
experiments (see Ref. \cite{heiblum} for an experimental realization) and assume perfect transmission probability in both branches (this implies $t_L=t_R$ and 
$\delta_\s=\pi/2$).
The conductance then reads 
\beq\label{gphi}
G=\frac{e^2}{h}(1+\cos(\vp/2)\sin(2 \pi e\Phi/h)).
\eeq
Two different periodicities appear with the orbital field $B_o$ responsible for the fluxes. 
First, a fast oscillation, of period $\frac{h}{eS}$, is mainly due to the flux in the section 
$S$ of the large loop. 
Second, a slow one comes from the flux in the small DD loop of area $s$. 
Its period is $\vp=4\pi$ instead of $2\pi$, owing to the 
spinor nature of the wavefunction. This results in slower (beating) oscillations, 
with period $\frac{2h}{es}$. 
Notice that the visibility of the fast AB oscillations is minima for 
a rotation of $\vp=2\pi$, for which the spinor changes its sign. 
This very striking consequence 
of quantum mechanics was proposed \cite{change-sign-theo} and verified \cite{change-sign-exp} with neutron interferometry. 
Let us for example compare this result to the situation in which the dots are
not capacitively coupled and are tuned independently in  the Kondo regime.
The conductance then reads $G=e^2/2h(1+\cos^2(\vp/2)+2\sin(2\pi e\Phi/h)\cos(\vp/2))$.
The conductance is also $4\pi$-periodic in $\vp$
but here the $4\pi$ period can be traced back to the spin-independent 
interference between the upper arm of the large ABI and both branches in
the lower arm. Nevertheless, this expression of the conductance is clearly different from the one in Eq. (\ref{gphi}) and precession can in principle be detected.
Another striking consequence of precession in branch II is
the spin polarization in the output, even when the incoming electrons are not polarized. Indeed
one finds that $\langle 2S_z \rangle=-\cos(\pi e(2\Phi)/h)\sin(\vp/2)$ while $\langle S^z\rangle=0$
without spin precession.
Maximum polarization comes from a destructive interference
occurring for one spin direction only. Testing the latter prediction requires spin filtering
only in the output and should be easier than the previous test based on Fig. 1.

For a closed interferometer, one may also try to compare the conductance in the large ABI obtained 
when the spin is precessing (with a SGI in the lower arm) to a reference case where no such precession is present (like for 
two independent quantum dots in the Kondo regime). We have described both forks of the large ABI by $3\times3$ S-matrices as in Ref. \cite{gefen}.
Nevertheless, one does not observe clear signatures through the conductance. Furthermore, the latter strongly depends on our choice of parameters entering  the S-matrices and we did not find any universal feature able to unambiguously distinguish between the two situations. It is therefore preferable to use an open large ABI.

Let us briefly discuss the feasibility of this proposal. 
Concerning the first experiment, one possibility is to use ferromagnetic semiconductors (Ga,Mn)As as lead
electrodes coupled to InAs quantum dots \cite{awschalom}.
Another promising route is to use carbon nanotube (CN)
quantum dots where Kondo effects have been shown \cite{kondocn}. They can be coupled to ferromagnetic electrodes and large magnetoresistance
effects have been observed recently \cite{takis}. 
Concerning the second experiment, two dots in  parallel
can be fabricated and inserted in an AB-loop
\cite{exptAB}, and a strong mutual capacitive coupling could be  achieved (the
residual tunneling amplitude  only needs to be smaller  than the Zeeman
energy\cite{apl}). This does not 
require any lead polarization if one searches for beating effects in the AB interferences.
This latter experimental proposal is achievable and may be easier than the first one.

In summary, we have shown how spin interferometry can be performed using orbital/spin entanglement in a 
double dot. The unitary transmission obtained in the Kondo regime is accompanied by spin 
precession, leading to a novel periodicity in an AB experiment. 
The authors acknowledge useful discussions with G. Zar\'and, P. Nozi\`eres, C.
Balseiro and T. Kontos.
This work was partially supported by  the contract PNano ``QuSpins'' of Agence Nationale de la Recherche.



\begin{thebibliography}{30}
\bibitem{equiv} Our set-up achieves the formal analogy noted in Ref. \onlinecite{konig}.
\end{thebibliography}
\end{document}